\documentstyle[twoside,fleqn,espcrc2]{article}

\newcommand{\AmS}{{\protect\the\textfont2
  A\kern-.1667em\lower.5ex\hbox{M}\kern-.125emS}}
\newcommand{\be}{\begin{equation}}
\newcommand{\ee}{\end{equation}}
\newcommand{\bea}{\begin{eqnarray}}
\newcommand{\eea}{\end{eqnarray}}
\newcommand{\ba}{\begin{array}}
\newcommand{\ea}{\end{array}}     
\newcommand{\nn}{\nonumber \\}
\newcommand{\half}{\frac{1}{2}}
\newcommand{\bb}{\bibitem}
\title{Gauge Theoretic Formulation of Dilatonic Gravity Coupled to
  Particles\thanks{
Talk presented at QG99 Meeting, Sardinia, September 1999.
This work is partially supported by CNPq.}}

\author{Victor O. Rivelles \\
{Universidade de S\~ao Paulo, Instituto de
  F\'\i sica \\ C. Postal 66318, 05315-970,  Sao Paulo, SP, Brazil \\
  e-mail: rivelles@fma.if.usp.br }%
        }
        
\begin{document}

\begin{abstract}
We discuss the formulation of the CGHS model in terms of a topological
BF theory coupled to particles carrying non-Abelian charge. 
\end{abstract}

\maketitle

In order to have a better understanding of the quantum properties of
four-dimensional gravity theories two-dimensional models have been
extensively studied. A particular model where black holes can
be formed and quantization can be performed is the 
Callan, Giddings, Harvey and Strominger (CGHS) model \cite{cghs}. It is
described by the action
\bea
\label{cghs}
S &=& \int d^2x \, \sqrt{-\bar{g}} [ e^{-2\phi}  ( \bar{R} - 4
\bar{g}^{\mu\nu} \partial_\mu \phi \partial_\nu \phi  - \Lambda )
\nn 
&+& \half \sum_i \bar{g}^{\mu\nu} \partial_\mu f_i \partial_\nu f_i
], 
\eea
where $\bar{R}$ is the curvature scalar build with the metric
$\bar{g}_{\mu\nu}$, $\phi$ is the dilaton, $\Lambda$ 
is the cosmological constant and $f_i$ is a set of scalar matter 
fields. If the conformal transformation 
$g_{\mu\nu} = e^{-2\phi} \bar{g}_{\mu\nu}$ 
is performed the action (\ref{cghs}) takes the form 
\be
\label{string}
S = \int d^2x \, \sqrt{-g}\, ( \eta R - \Lambda  + \half \sum_i
{g}^{\mu\nu} \partial_\mu f_i \partial_\nu f_i ), 
\ee
where $\eta = e^{-2\phi}$ and $R$ is the scalar curvature built with the
metric $g_{\mu\nu}$. Now the field equation for $\eta$ implies that $R
= 0$ so that the two-dimensional space-time is locally flat and there
is no black hole solution. While the model described by (\ref{cghs}) presents
Hawking radiation the model described by (\ref{string}) has no Hawking
radiation. It has been argued that with proper care of the conformal
transformation no ambiguity exists  \cite{cadoni}.
Even so the quantization in either form is not free of troubles
\cite{jackiw1}.

An important feature of the action (\ref{string}) is that it can be
cast as a topological gauge theory of the BF type 
with a gauge group which is a central extension of the two-dimensional
Poincar\'e group  \cite{cangemi}. If matter is coupled in this
formulation it should be coupled in a gauge invariant way.
A possibility makes use of a formulation of
relativistic particles which carry non-Abelian charges \cite{bal}. It
was applied to the gauge theoretic version of dilatonic gravity
\cite{stephany}. An important aspect is that the curvature equation
$R=0$ never acquires a 
source term. We will show that in the formulation where particles
carry non-Abelian charge a new gauge invariant coupling does exist for
topological BF theories. This new
coupling provides a source term for the curvature equation and black
hole solutions can then be found \cite{us}.

Particles carrying non-Abelian degrees of freedom  were originally
introduced in the context of QCD \cite{wong,bal}. They are described by
the group element $g(\tau)$ and a real constant element of the algebra
$K$, $\tau$ being the proper time of the particle. It is useful to
introduce the variable 
\be
\label{Q}
Q(\tau) = g(\tau) K g^{-1}(\tau),
\ee
which is in the adjoint representation. As usual a covariant
derivative can be introduced
\be
\label{cov_derivative}
D_\tau = \frac{d}{d\tau} + e \dot{x}^\mu A_\mu( x(\tau) ). 
\ee
If we also consider a kinetic term for the relativistic particle then
an action which is gauge and reparametrization invariant is \cite{bal}
\be
\label{bal}
S = -m \int d\tau \, \sqrt{\dot{x}^2} + \int d\tau \, Tr( K
g^{-1}(\tau) D_\tau g(\tau) ).
\ee
This action is also invariant under the
transformation $K \rightarrow S K S^{-1}$ where $S$ is $\tau$
independent. This shows that the action (\ref{bal}) is independent of
the direction in the internal symmetry space 
given by $K$.
Varying the action (\ref{bal}) with respect to $x^\mu(\tau)$ we get a
non-Abelian version of the Lorentz force
\be
\label{lorentz}
m \ddot{x}^\mu + \Gamma^\mu_{\nu\rho} \dot{x}^\nu \dot{x}^\rho = -e Tr(
F^\mu_{\,\,\,\,\nu} Q) \dot{x}^\nu,
\ee
while varying with respect to $g(\tau)$ we get a covariant conservation
equation for the non-Abelian charge $Q$
\be
\label{conservation}
\frac{dQ}{d\tau} + [ A_\mu(x(\tau)),
Q(\tau) ]   \dot{x}^\mu (\tau) = 0.
\ee
These equations are known as the Wong equations \cite{wong}. 

Consider now a two-dimensional BF topological field theory
\be
\label{bf}
S = \int d^2x \, Tr (\eta \, F),  
\ee
where $F = dA + A^2$ is the curvature two-form corresponding to the
connection one-form $A$ and $\eta$ is a zero-form transforming in the
co-adjoint representation of the gauge group.
Since the structure of the BF theory requires two fields then, besides
the coupling involving the gauge field $A$, we can
consider another coupling involving the Lagrange multiplier 
$\eta$. A coupling of the type $Tr( \eta \, Q)$  is gauge invariant
but not  
proper time reparametrization invariant. In order to get a
reparametrization invariant action we introduce the worldline einbein
$e(\tau)$ and the respective mass term for the particle. So we can
consider an extension of the former actions to
\bea
\label{new}
&S& = \int d^2x \, Tr( \eta \, F) + \int d\tau \, Tr(g^{-1} K
\dot{g} )  \nn
&+& e \int d^2 x \int d\tau   Tr( Q(\tau) A_\mu (x)) \,
\delta^2(x - x(\tau)) \, \dot{x}^\mu   \nn
&+& g \int d^2x \int d\tau   e(\tau) \, 
Tr( Q(\tau) \eta) \, \delta^2( x - x(\tau)) \nn &+& \half m^2 \int d\tau \,
e(\tau), 
\eea
where $e$ and $g$ are independent coupling constants.
The action (\ref{new}) is invariant under gauge transformations
\bea
\label{gauge-transf}
A &\rightarrow& h A h^{-1} - dh h^{-1}, \qquad
\eta \rightarrow h \eta h^{-1}, \nn
g &\rightarrow& h g, \qquad
K \rightarrow K,
\eea
and proper time reparametrization
\be
\label{proper-time}
e^\prime(\tau^\prime) = \frac{d\tau}{d\tau^\prime} e(\tau), \qquad
\dot{x}^{\prime\mu}(\tau^\prime) = \frac{d\tau}{d\tau^\prime}
\dot{x}^\mu(\tau), 
\ee
with all remaining fields being reparametrization scalars. 

The new coupling now implies that the connection is no longer flat and
has as source the non-Abelian charge $Q$
\be
\label{f=q}
\epsilon^{\mu\nu} F_{\mu\nu} = g \int d\tau \, e(\tau) \,Q(\tau) \,
\delta^2(x - x(\tau)). 
\ee
This will allows us to find  black hole solutions \cite{us}. 
The field equation for the Lagrange multiplier
is
\be
\label{deta}
\epsilon^{\mu\nu} D_\nu \eta = e \int d\tau \, Q(\tau) \, \delta^2
(x - x(\tau)) \, \dot{x}^\mu(\tau),
\ee
while the equation for non-Abelian charge is modified to 
\be
\label{dq}
\frac{dQ}{d\tau} + e [ A_\mu (x(\tau)), Q ] \dot{x}^\mu + g 
[ \eta (x(\tau)), Q] e(\tau) = 0,
\ee
generalizing the conservation equation (\ref{conservation}) to the BF
theory. The field equation obtained by varying the worldline einbein
is
\be
\label{einbein}
g \int d^2x \,\, Tr(Q(\tau) \eta(x) ) \, \delta^2(x - x(\tau) ) +
\half m^2 = 0.
\ee

For the two dimensional dilatonic gravity theory we choose
the gauge group as the central extension of the Poincar\'e
group \cite{cangemi}
\bea
\label{group}
{}[P_a, P_b] &=& \epsilon_{ab} Z, \qquad
{}[J, P_a] = {\epsilon_a}^b P_b, \nn
{}[P_a, Z] &=& [J, Z] = 0, 
\eea
where $P_a$ is the translation generator, $J$ is the Lorentz transformation 
generator and $Z$ is a central element of the group. The
supersymmetric extension of 
(\ref{group}) was performed in \cite{me}. When we consider the algebra
(\ref{group}) we can expand the one form gauge potential  
in terms of the generators of the algebra
\be
\label{potential}
A = e^a P_a + w J  + A Z.
\ee
The fields $e^a, w$ and $A$ are going to be identified with the 
zweibein, the spin connection and an Abelian gauge field, respectively. 
The Lagrange multiplier $\eta$ can be expanded as
\be
\label{eta}
\eta = \eta^a P_a + \eta_3 J + \eta_2 Z,
\ee
with components $\eta^a, \eta_2$ and $\eta_3$ with $\eta_2$ being
proportional to the dilaton in  (\ref{string}). 
Then the curvature two-form $F$ has components
\bea
\label{f}
F^a(P) &=& de^{a} + we^{b}\epsilon_{b}^{\,\,a}, \nn
F(J) &=& d\omega, \nn
F(Z) &=& dA + \half e^a e^b \epsilon_{ab}.
\eea
Similarly the non-Abelian charge $Q$ can be expanded as 
\be
\label{q}
Q = Q^a P_a + Q_3 J + Q_2 Z.
\ee

The field equations for the gauge fields (\ref{f=q}) are 
\bea
\label{f1}
& &\epsilon^{\mu\nu} ( \partial_\mu e_\nu^a + \omega_\mu e_\nu^b
\epsilon_b^{\,\,a} ) \nn &+& g \int d\tau \, e(\tau) \, Q^a(\tau) \,
\delta^2(x - x(\tau)) = 0, \\
\label{f2}
& &\epsilon^{\mu\nu} \partial_\mu \omega_\nu \nn &+& g \int d\tau \,
e(\tau) \, Q_3(\tau) \, \delta^2(x - x(\tau))  = 0, \\
\label{f3}
& &\epsilon^{\mu\nu} (\partial_\mu a_\nu + \half e_\mu^a e_\nu^b
\epsilon_{ab}) \nn &+& g \int d\tau \, e(\tau) \, Q_2(\tau) \, \delta^2(x -
x(\tau)) = 0, 
\eea
while the field equations for the Lagrange multipliers (\ref{deta}) are
\bea
\label{eta2}
& &\epsilon^{\mu\nu} ( \partial_\nu \eta_a - \omega_\nu
\epsilon_a^{\,\,b} \eta_b + \eta_3 \epsilon_{ab} e_\nu^b ) \nn 
&+&  e \int d\tau Q_a(\tau) \delta^2(x - x(\tau)) \dot{x}^\mu 
(\tau) = 0, \\
\label{eta1}
& &\epsilon^{\mu\nu} ( \partial_\nu \eta_2 + e_\nu^a \epsilon_a^{\,\,b}
\eta_b ) \nn &+& e \int d\tau \,\, Q_2(\tau) \,\, \delta^2(x - x(\tau))
\dot{x}^\mu (\tau) =  0, \\ 
\label{eta3}
& &\epsilon^{\mu\nu} \partial_\nu \eta_3 \nn &+&  e \int d\tau \,\,
Q_3(\tau) \,\, \delta^2(x - x(\tau)) \dot{x}^\mu (\tau) =  0.
\eea
The equations of motion for the non-Abelian charge (\ref{dq}) are
\bea
\label{q1}
& &\frac{dQ^a}{d\tau} + e \epsilon_b^{\,\,a} (e_\mu^b  
Q_3 - \omega_\mu Q^b ) \dot{x}^\mu \nn &+& g \epsilon_b^{\,\,a} (\eta^b Q_3
- \eta_3 Q^b ) e(\tau) = 0, \\
\label{q3}
& &\frac{dQ_2}{d\tau} - \epsilon_{ab} (e e_\mu^a
Q^b \dot{x}^\mu - g \eta^a Q^b e(\tau) ) = 0, \\
\label{q2}
& &\frac{dQ_3}{d\tau} = 0.
\eea
Finally the equation of motion (\ref{einbein}) gives a constraint
among the 
non-Abelian charge, the Lagrange multiplier on the worldline and the
particle mass
\bea
\label{einbein2}
&& g \int d^2x \, ( Q^a \eta_a + Q_3 \eta_2 + Q_2 \eta_3 ) \delta^2(x -
x(\tau)) \nn &+& \half m^2 = 0. 
\eea

We will discuss some solutions of the above equations. 
In order to solve (\ref{f1}-\ref{einbein2}) we have to perform
several gauge fixings. We also have to choose a space-time trajectory
for the 
particle. We will look for static solutions so we use Rindler like 
coordinates $(x,t)$. For simplicity let us consider the proper time
gauge for the particle $e(\tau) = 1$ and set the particle at
rest in the origin $x(\tau) = 0, \, t(\tau) = \tau$.

Let us consider first the gauge field sector. In the
gravitational sector we will choose a diagonal zweibein $e_0^1 = e_1^0
= 0$ with the non vanishing components satisfying $e_0^0 = ( e_1^1
)^{-1}$, 
and a vanishing space component of the connection $\omega_1 = 0$. For
the Abelian gauge field we will choose the axial gauge $A_1 = 0$.
eqs.(\ref{f1}-\ref{f3}) reduce to
\bea
\label{f_est1}
\partial_1 e_0^0 + \omega_0 e_1^1 &=& g Q^0 \, \delta(x), \\
\label{f_est2}
0 &=& g Q^1 \delta(x), \\
\label{f_est3}
\partial_1 \omega_0 &=& g Q_3 \, \delta(x), \\
\label{f_est4}
\partial_1 A_0 + e_0^0 e_1^1 &=& g Q_2 \, \delta(x).
\eea

If no matter is present ($ e = g = 0 $) then we find flat
space-time as the only solution.
Now consider the situation when matter is present. If $Q_3 = 0$ and
$Q^a \not= 0$ then the space-time has torsion but no curvature. If
$Q_3 \not= 0$ and $Q^a = 0$ then the space-time has curvature but
no torsion. Let us consider the last case.  Take $Q_2$ and $Q_3$ as 
constants (as we shall see below $Q_2$ and $Q_3$ constants and $Q^a =
0$ is a solution of (\ref{q1}-\ref{q2}) ). Then we find as solution
of (\ref{f_est1}-\ref{f_est4}) 
\bea 
\label{schw}
\omega_0 &=& g Q_3 \epsilon(x), \quad e_0^0 = ( \tilde{b} - 2 g 
Q_3 |x| )^\half, \nn
A_0 &=& - x + g Q_2 \epsilon(x) + \tilde{A},
\eea
where $\tilde{b}$ and $\tilde{A}$ are integration constants. The
space-time described 
by (\ref{schw}) has a black hole and the curvature scalar
is given by (\ref{f_est3}) $R = g Q_3 \delta(x)$. Notice that $g Q_3$
can now be understood as the black hole mass and it is essential to
have $g \not= 0$. 

For the Lagrange multiplier sector the gauge choices reduce
eqs.(\ref{eta2}-\ref{eta3}) to
\bea
\label{eta_est}
& & \partial_1 \eta_0 - \eta_3 e_1^1 = -e Q_0 \delta(x), \nn
& & \partial_1 \eta_1 = -e Q_1 \delta(x), \nn
& & \omega_0 \eta_1 = \omega_0 \eta_0 + \eta_3 e_0^0 = 0, \nn
& & \partial_1 \eta_2 - e_1^1 \eta_0 = -e Q_2 \delta(x), \nn
& & e_0^0 \eta_1 = 0, \nn
& &\partial_1 \eta_3 = -e Q_3 \delta(x).
\eea
In the presence of matter with $Q_2$ and $Q_3$ constants and $Q^a = 0$
we find, using (\ref{schw}) 
\bea
\label{eta_schw}
\eta_0 &=& \frac{e}{g} (\tilde{b} - 2 g Q_3 |x|)^\half, \quad \eta_1 =
0, \nn 
\eta_2 &=& \frac{e}{g} x - e Q_2 \epsilon(x) + \tilde{c}, \quad
\eta_3 = - e Q_3 \epsilon(x), 
\eea
where $\tilde{c}$ is another integration constant and $\epsilon(x)$ is
the step function. The appearance of the step function in the solution
for the Lagrange multiplier fields signals that there are topological
restrictions to the motion of particles.  It is
remarkable that the would be cosmological constant $\eta_3$ is now a
step function changing sign at the position of the particle.  The
dilaton $\eta_2$ still has its linear term but has also acquired a
step function. We can however set $Q_2=0$ and still have a linear
dilaton and the black hole (\ref{schw}), which is independent of
$Q_2$. 

For the non-Abelian charge the equations 
(\ref{q1}-\ref{q2}), after the gauge choice, reduce to
\bea
\label{q_est}
& &\dot{Q}^0 - e Q^1 \omega_0 - g ( Q^1 \eta_3 - Q_3 \eta^1
) = 0, \nn
& &\dot{Q}^1 - e ( Q^0 \omega_0 - Q_2 e_0^0 ) - g ( Q^0
\eta_3 - Q_3 \eta^0 ) = 0, \nn
& &\dot{Q}_2 - e Q^1 e_0^0 + g ( Q^0 \eta^1 - Q^1 \eta^0 ) = 0, \nn
& &\dot{Q}_3 = 0.
\eea
In the presence of matter with $Q^a = 0$ and $Q_2$ and $Q_3$ constants
we find using (\ref{schw}) and  
(\ref{eta_schw}) that $Q_2$ and $Q_3$ are constants as we had
anticipated. 

Finally the constraint equation (\ref{einbein2}) becomes, after the
gauge choices, 
\bea
\label{einbein3}
&&g [ Q^a(\tau) \eta_a(\tau,0) + Q_3(\tau) \eta_2(\tau,0) \nn  &+& Q_2(\tau)
\eta_3(\tau,0) ] + \half m^2 = 0.
\eea
In the presence
of matter with $Q^a = 0$ (\ref{einbein3}) is ill defined since
according to (\ref{eta_schw}) $\eta_2$ and $\eta_3$ have a
discontinuity at $x=0$. We then take $Q_2 = 0$ and (\ref{einbein3})
becomes $g Q_3 \tilde{c} + \half m^2 = 0$ giving a constraint among
the integration constants $Q_3$ and 
$\tilde{c}$ and the mass $m$. Recalling that the curvature scalar is
$R = g Q_3 \delta(x)$ we can interpret the black hole mass as being
due to the non-Abelian charge of the particle $Q_3$ or to its mass
$m$. 

Some local solutions for the gauge theoretic version
of dilatonic gravity theories with non-Abelian sources have been
presented. Presently we are investigating the constraint structure of
the model and its Hamiltonian formulation. We also plan to study
global aspects of the model.

\end{document}